\documentclass[aps,prl,twocolumn,showpacs,floats]{revtex4}
\usepackage{bm}
\usepackage{graphicx}
\newcommand{\bi}[1]{{\boldsymbol{#1}}}

\begin{document}

\title{Hidden-Symmetry-Protected Topological Semimetals on  a Square Lattice  }
\author{Jing-Min Hou}\email{jmhou@seu.edu.cn}
\affiliation{Department of Physics, Southeast University, Nanjing,
211189, China}
\begin{abstract}
We study a    two-dimensional fermionic
square lattice, which supports the existence of two-dimensional Weyl semimetal, quantum anomalous Hall effect, and $2\pi$-flux topological semimetal in different parameter ranges.       We show that the band degenerate points of the two-dimensional Weyl semimetal and $2\pi$-flux
topological semimetal are  protected by two distinct novel hidden symmetries, which both corresponds to antiunitary composite operations. When these hidden symmetries are broken, a gap opens between the conduction and valence bands, turning the system into a insulator. With appropriate parameters, a quantum anomalous Hall effect emerges. The degenerate point  at the boundary between the quantum anomalous Hall insulator and trivial band insulator is also protected by the hidden symmetry.


\end{abstract}
\maketitle

 \em Introduction.\em ---The research on topological phases is becoming an increasingly
 important theme in condensed
matter physics.  Topological matters are classified according to
topological invariants rather than symmetries\cite{Hasan,Qi}.
Depending on the dimensionality and the symmetry classes specified
by time reversal symmetry and particle-hole symmetry,   gapped
systems can be classified into ten types of topological
phases\cite{Schnyder}, such as the integer quantum Hall
states\cite{Thouless}, quantum anomalous Hall insulator\cite{Haldane,LiuQA}, topological insulators\cite{Kane}, chiral
topological superfluids\cite{Read}, and helical topological
superfluids or superconductors\cite{Qi2}.
 More recently, physicists have found that,
 besides the gapped systems,   gapless systems can also
support topological phases, i.e., topological
semimetals\cite{Wan,Xu,Burkov1,Burkov2,Fang,Zyuzin,Sun,Jiang,Hosur,Delplace}.

Topological semimetals have
band structures with band-touching points in momentum space, where the isolated band degeneracy occurs. In general, at these kind of band-degeneracy points, there exist singularities of a Berry field. Around
the singularities,  vortex
structures in two-dimensions or monopoles in three dimensions appear.
These vortices or monopoles with opposite topological charges  are separated from each other in
momentum space due to symmetries and thus  cannot be destroyed   by the mutual
annihilation of pairs with opposite topological charges.
 In three dimensions, the band degeneracy at isolated points can be
accidental \cite{Wan, Herring}. The robustness
of the accidental degeneracy depends on its
codimension\cite{Burkov1}. However, such accidental band degeneracies are vanishingly improbable in one
and two dimensions if there are not additional symmetry
constraints\cite{Balents}.
Therefore, in two dimensions, the band
degeneracy at isolated points must be protected by symmetries.

\begin{figure}[ht]
\includegraphics[width=0.46\columnwidth]{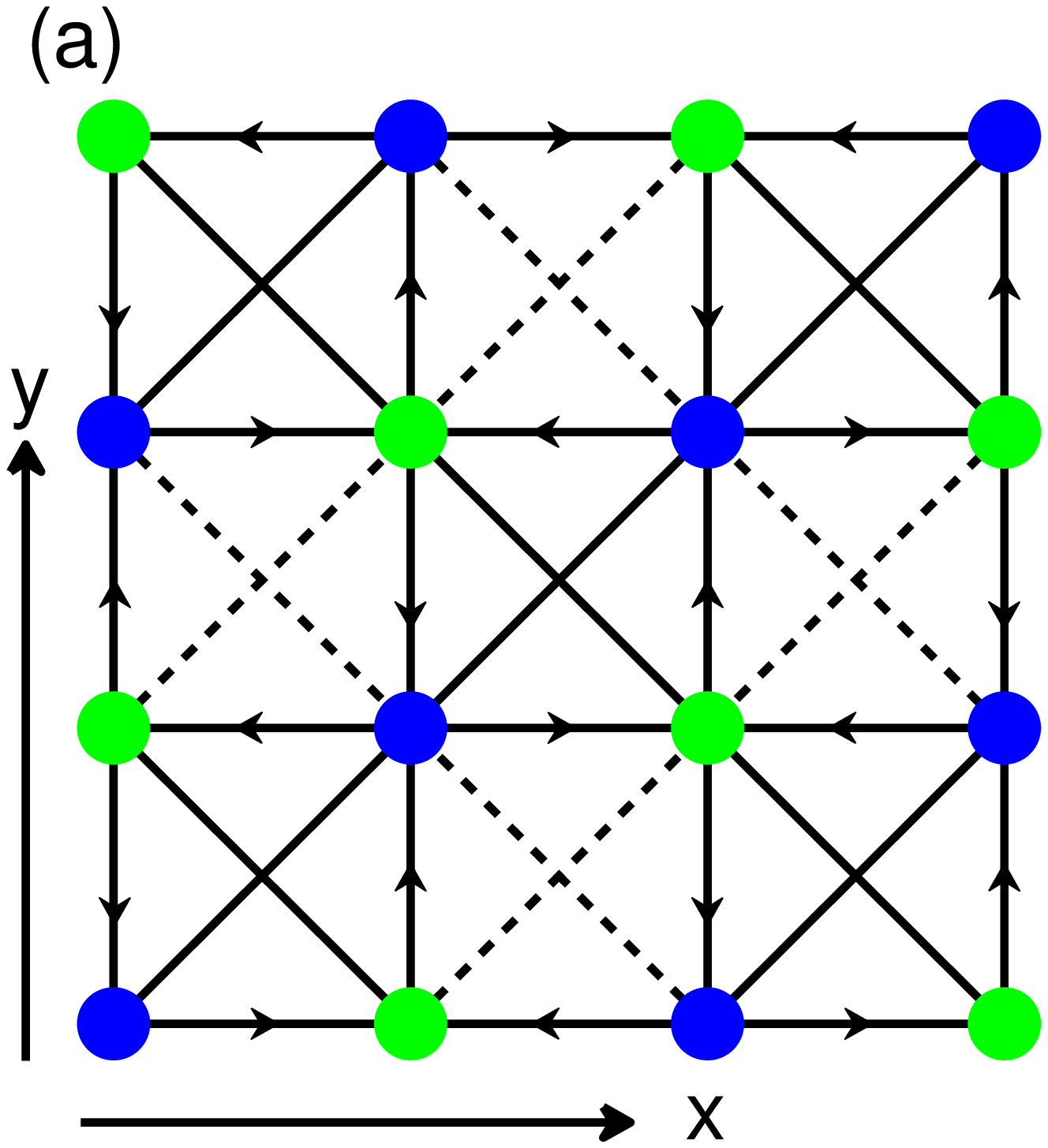}
\includegraphics[width=0.52\columnwidth]{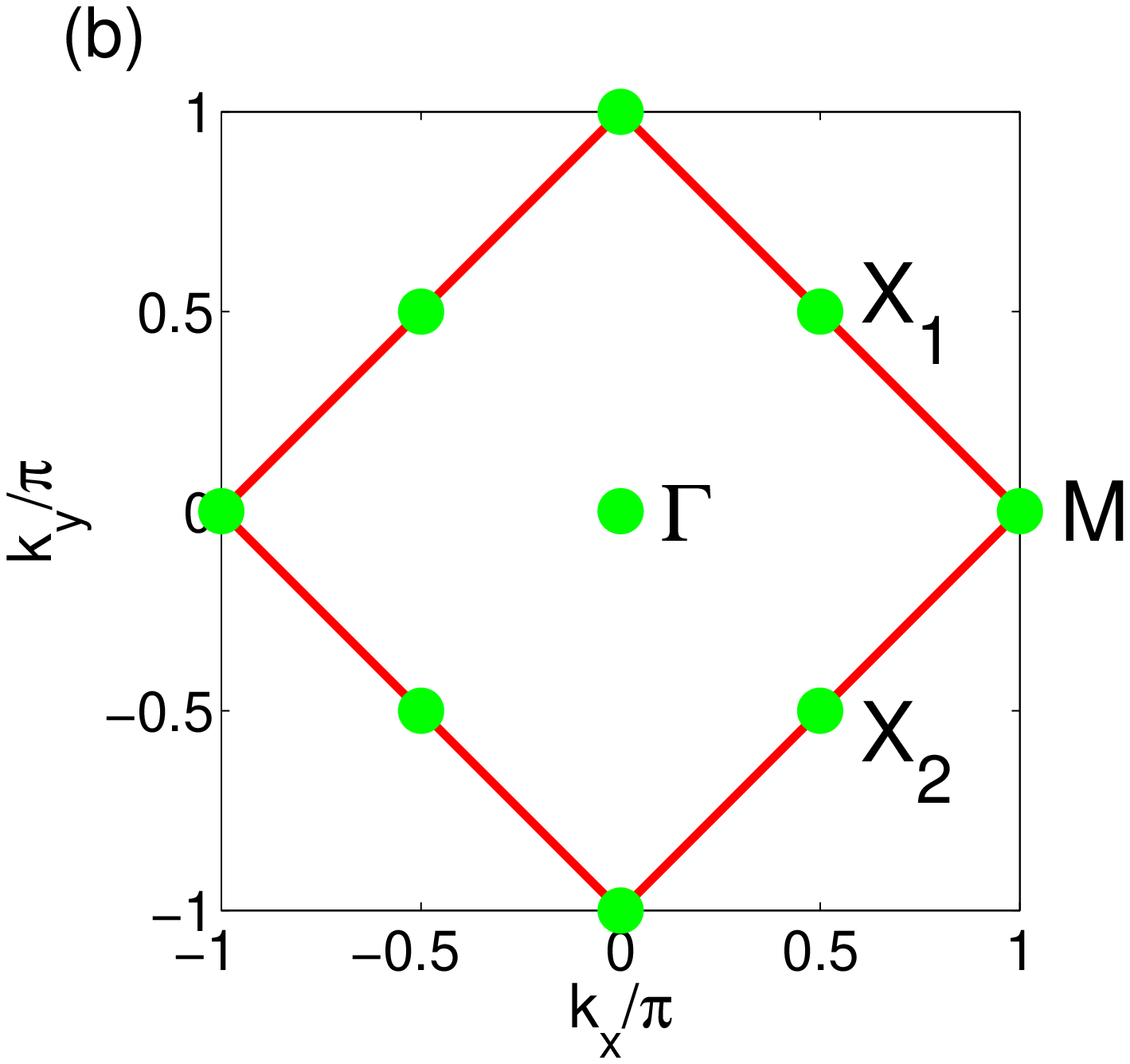}
\caption{(Color online).  (a)  Schematic of the lattices. The arrows represent the accompanying phase of hopping $\gamma$ and the dashed lines indicate a $\pi$ accompanying phase;  the blue  and green
filled circles represent the lattice sites of sublattices $A$ and $B$,
respectively. (b) The first Brillouin zone, which is surrounded by red lines. Here, the green filled circles represent symmetry points denoted by $\Gamma$, $M$ and $X_{1,2}$, respectively. }\label{fig1}
\end{figure}

Discrete symmetries with  antiunitary operators play a crucial role in topological phases, for instance, time-reversal symmetry and particle-hole symmetry
offer a base for classifying the topological phases into ten classes\cite{Schnyder}. In condensed matter physics, besides time-reversal and particle-hole symmetries, there exists a class of hidden symmetries, which are seldom studied in the literature. These hidden symmetries are discrete symmetries with antiunitary composite operators, which, in general,  consist  of  translation, complex conjugation and sublattice exchange, and sometimes also include local gauge transformation and rotation. Although a lot of researches were done focusing  on  Dirac or Weyl semimetals in two-dimensional lattices in recent years\cite{Lim,Hou,Goldman,Bercioux,Goldman2}, few of them have involved   the relation between the band degeneracies  and the above hidden symmetries.
  In this Letter, we will study this relation and take a square lattice as an example. This
  square lattice  supports the existence of topological semimetals and a quantum anomalous Hall effect in different parameter ranges.
The topological semimetals appearing in this model include a two-dimensional Weyl semimetal and a $2\pi$-flux topological semimetal, which are protected by different hidden symmetries. Moreover, we  show that the Weyl node at the phase boundary between quantum anomalous Hall insulator and trivial band insulator is protected by the hidden symmetry as well.
  In order to show the ubiquity of
 hidden-symmetry-protected topological semimetals,  we also present other two-dimensional lattices with different hidden symmetries in the Supplemental Material\cite{SM}.

\em Model.\em---We consider a square lattice as shown in Fig.\ref{fig1}. Because of the presence of the accompanying phases of hopping, the translation symmetry is broken, then the lattice is divided into  two sublattices denoted $A$  and
$B$.  This square lattice has a
lattice spacing  $d$ and,   for simplicity, we assume $d=1$ in the following process. For each sublattice, the primitive lattice vectors are
defined as $\bi{a}_1=(1,1)$ and $\bi{a}_2=(1,-1)$. For the reciprocal lattice,  the corresponding primitive
reciprocal lattice vectors are $\bi{b}_1= (\pi,\pi)$
and $\bi{b}_2=(\pi,-\pi)$.

The corresponding total Hamiltonian consists of three parts such as the square lattice Hamiltonian $H_0$, the diagonal Hamiltonian $H_1$ and the staggered potential Hamiltonian $H_2$, i.e., $H=H_0+H_1+H_2$,  which can be written as,
\begin{eqnarray}
  H_0&=&-t\sum_{i\in A} [ e^{-i\gamma} a^\dag_{i}
b_{i+\hat{x}}  +  e^{-i\gamma} a^\dag_{i} b_{i-\hat{x}}
  +  e^{ i\gamma} a^\dag_{i} b_{i+\hat{y}}\nonumber\\
  && + e^{ i\gamma}a^\dag_{i}b_{i-\hat{y}}]+ {\rm
H.c.},\label{H0}
\end{eqnarray}
and
\begin{eqnarray}
  H_1&=&-t_1\sum_{i\in A} [  a^\dag_{i}
a_{i+\hat{x}+\hat{y}}  -a^\dag_{i}
a_{i+\hat{x}-\hat{y}}]\nonumber\\
&&  +t_1\sum_{i\in B} [ b^\dag_{i} b_{i+\hat{x}+\hat{y}}  -b^\dag_{i} b_{i+\hat{x}-\hat{y}}
 ]+{\rm H.c.},\label{H1}
\end{eqnarray}
and
\begin{eqnarray}
H_2=v \sum_{i\in A} a^\dag_ia_i-v\sum_{j\in B}b^\dag_jb_j,\label{H2}
\end{eqnarray}
where  $a_i  $   destructs  a
particle at the site  $i$ in
sublattice $A$ and  $b_j  $
destructs  a particle   at the site $j$ in
sublattice $B$;  $\hat{x}\equiv(1,0)$ and
$\hat{y}\equiv (0,1)$ represent the unit vectors in the $x$ and $y$ directions, respectively; $0<\gamma<\pi/2$ is a phase factor along
with hopping; $t$ is the hopping amplitude along the
$x$ and $y$ directions and $t_1$ the hopping amplitude along the diagonal directions; $v$ is the magnitude of staggered potential.
In the following process, we will consider the model in three parameter ranges such as (i) $0<\gamma<\pi/2$, $t\neq 0$, $t_1=0$, $v=0$, i.e. $H=H_0$,
(ii) $0<\gamma<\pi/2$, $t\neq 0$, and at least one of $t_1$ and $v$ being nonzero,    and (iii) $\gamma=0$, $t\neq 0$, $t_1\neq 0$, $v=0$, which support the existence of a two-dimensional Weyl semimetal, quantum anomalous Hall effect, and $2\pi$-flux
topological semimetal, respectively.

\em Two-dimensional Weyl semimetals.\em---First, we consider the first parameter range with $0<\gamma<\pi/2$, $t\neq 0$, $t_1=0$ and $v=0$, i.e. $H=H_0$.
 We take Fourier transformation to annihilation operators  as $
{a}_{\bi{k}}=\frac{1}{\sqrt{N}}\sum_{i}
  {a}_{i}e^{-i\bi{k}\cdot\bi{R}^A_i} $, ${b}_{\bi{k}}=\frac{1}{\sqrt{N}}\sum_{i }
  {b}_{i}e^{-i\bi{k}\cdot\bi{R}^B_i} $\cite{Bena} and   define the two-component annihilation operator as $
 {\eta}_{\bi{k}}\equiv\frac{1}{\sqrt{2}}[
 {a}_{\bi{k}}, {b}_{\bi{k}}]^T$.
 The Hamiltonian
(\ref{H0}) can be rewritten as $ H=\sum_{\bi{k}}
 {\eta}_{\bi{k}}^\dag{\cal H}(\bi{k})
  {\eta}_{\bi{k}}
$ with
\begin{eqnarray}
{\cal H}(\bi{k})&=&- \cos\gamma \Omega_+\sigma_x - \sin\gamma \Omega_-\sigma_y,
\label{BH0}
\end{eqnarray}
where $\Omega_+=2t(\cos k_x+\cos k_y)$ and $\Omega_-=2t(\cos k_x -\cos k_y)$; $\sigma_x$ and $\sigma_y$ are the Pauli matrices.  Diagonalizing Eq.(\ref{BH0}), we obtain the dispersion relation as
$E({\bi{k}}) =
  \pm   \sqrt{\cos^2\gamma \Omega_+^2+\sin^2\gamma \Omega_-^2}$.

\begin{figure}[ht]
\includegraphics[width=0.46\columnwidth]{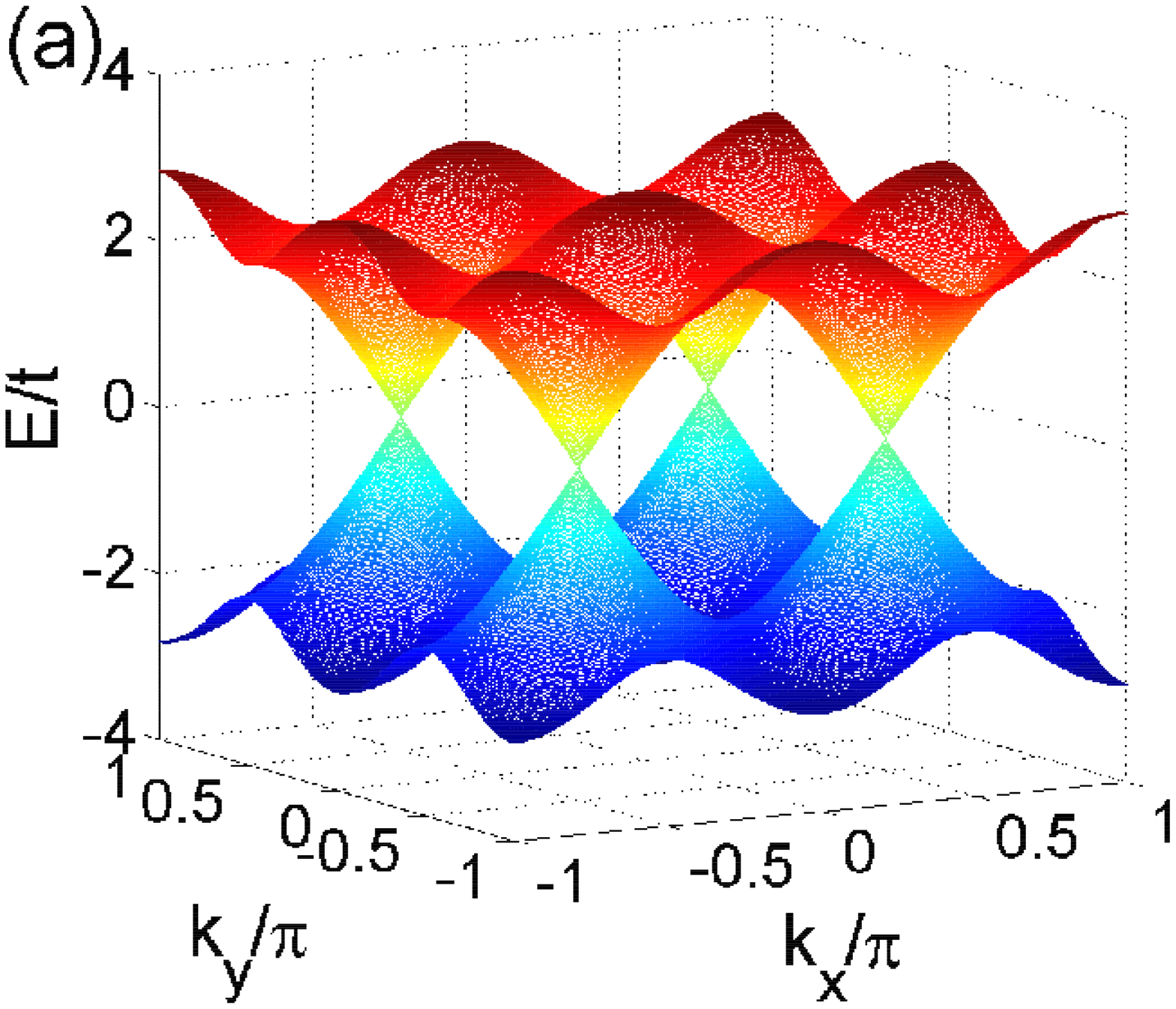}
\includegraphics[width=0.46\columnwidth]{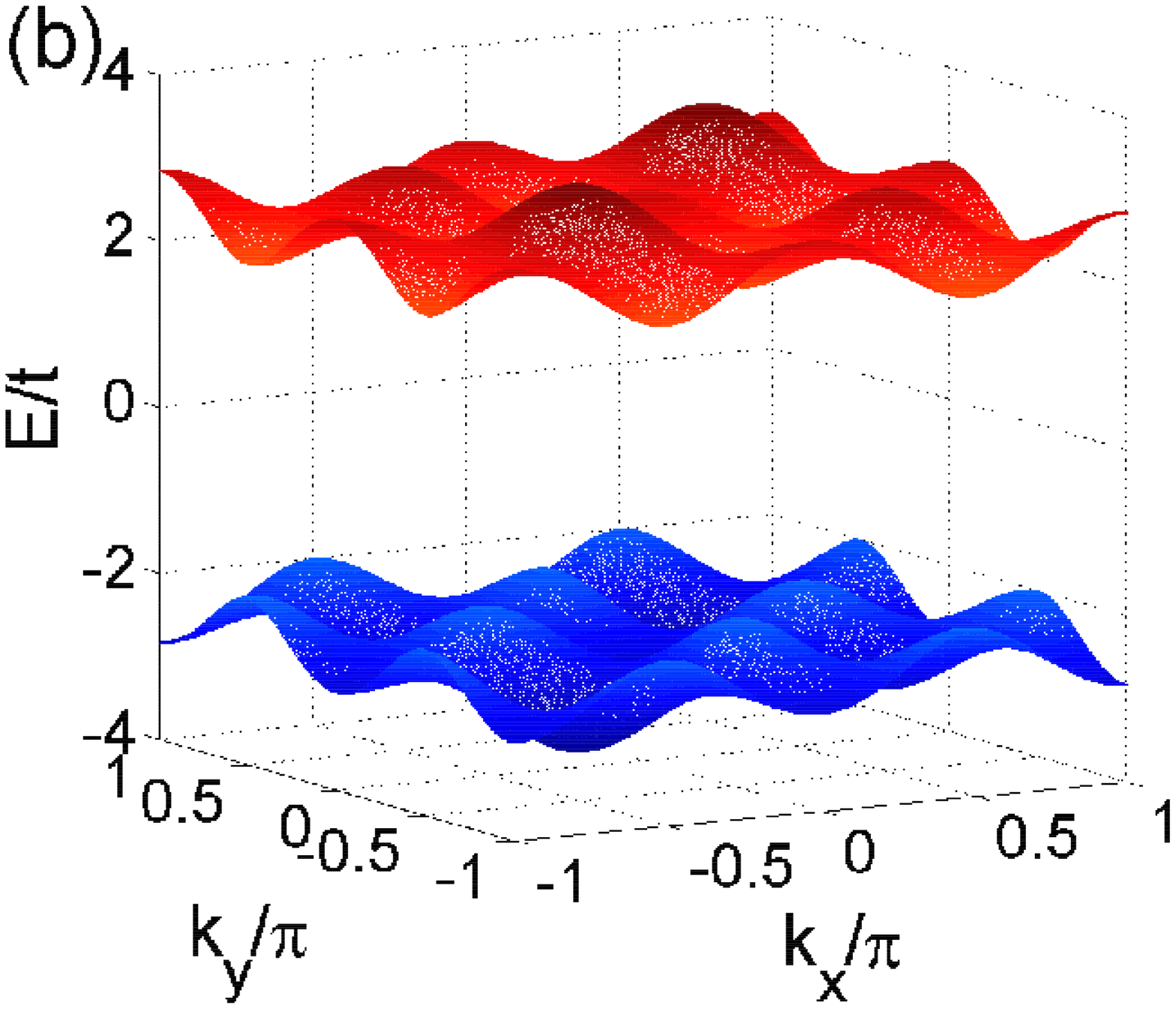}
\includegraphics[width=0.46\columnwidth]{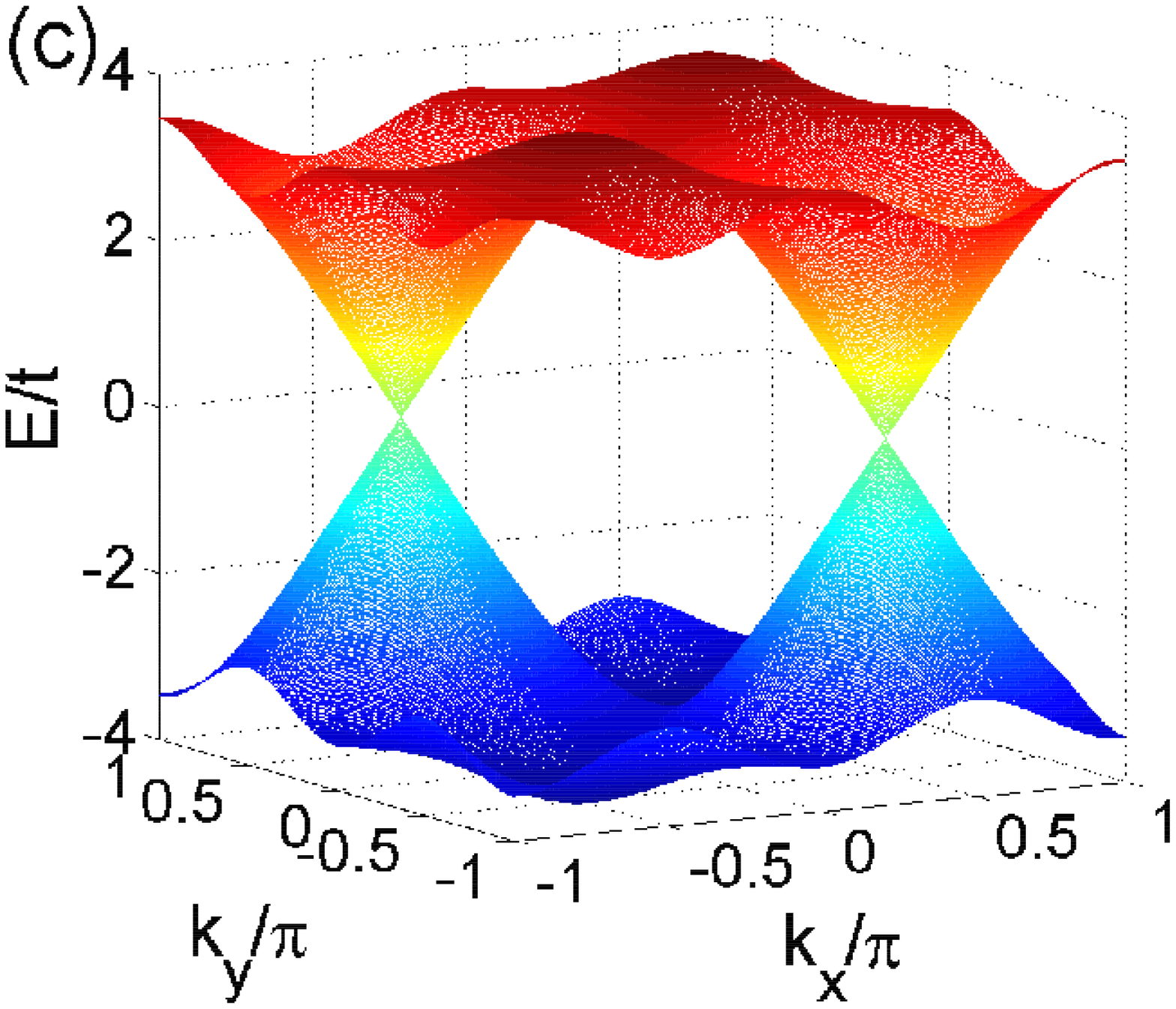}
\includegraphics[width=0.46\columnwidth]{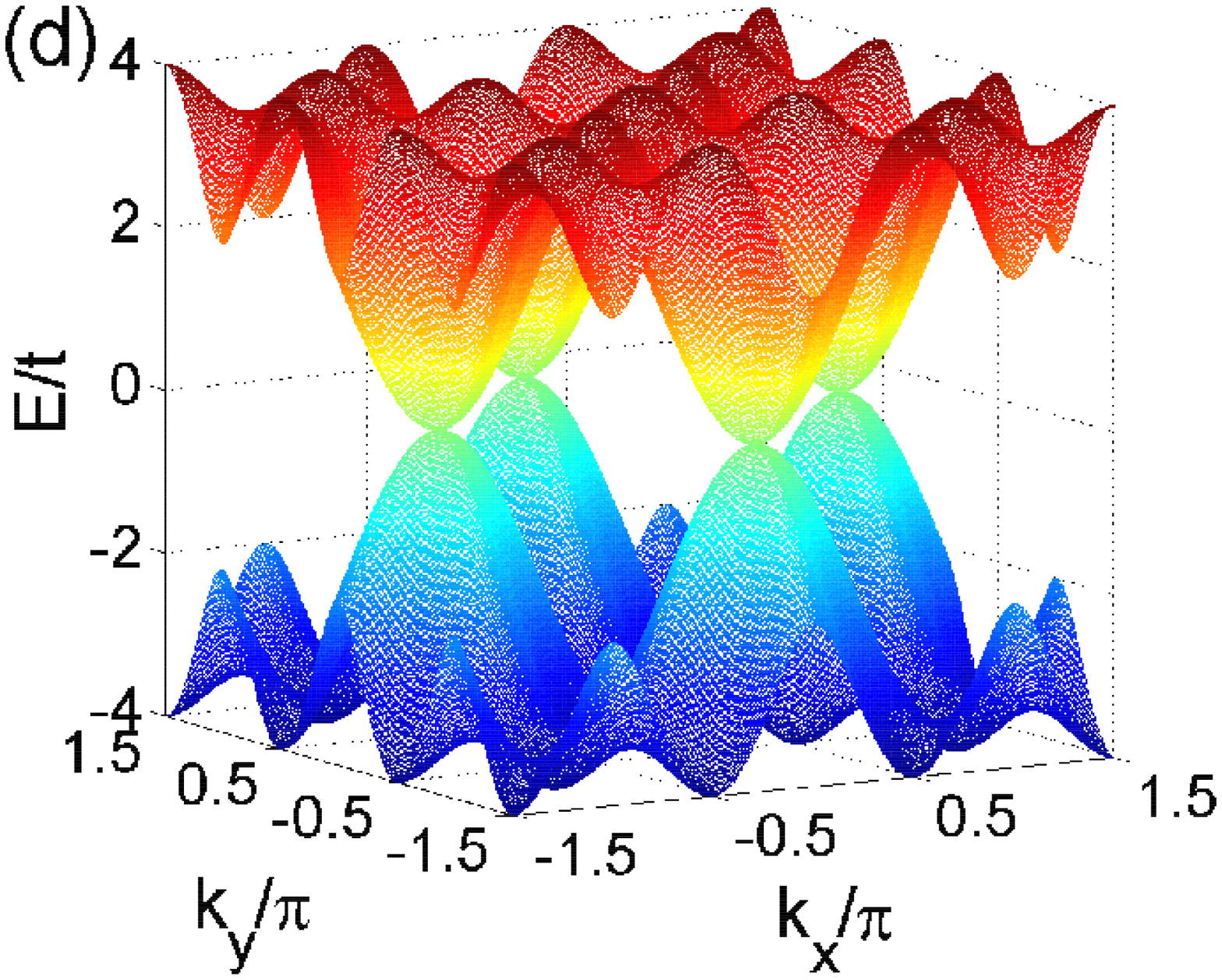}
\caption{(Color online).    The energy dispersions
 for (a)a two-dimensional Weyl semimetal with $\gamma=\pi/4$, $t_1=0$, $v=0$; (b)a quantum anomalous Hall insulator with $\gamma=\pi/4$, $t_1=0.8t$, $v=0.2t$;  (c) the boundary between the quantum anomalous Hall insulator and trivial band insulator with $\gamma=\pi/4$, $t_1=0.5t$, $v=2t$; (d)a $2\pi$-flux topological semimetal with $\gamma=0$, $t_1=t$ and $v=0$.
}\label{fig2}
\end{figure}

Figure \ref{fig2}(a) shows the dispersion relation for  $\gamma=\pi/4$.
From the dispersion relation, we find that
the conduction   and  valence bands are touched at symmetry points $X_{1,2}=(\pi/2,\pm\pi/2)$ in the Brillouin zone as denoted in Fig.\ref{fig1}(b). Near these
degenerate points, the band structure has a conelike shape and
  is linear. Thus,  the
quasiparticles and quasiholes behave like   massless relativistic fermions. Around these degenerate points, the single-particle Hamiltonian (\ref{BH0}) can be
linearized as
\begin{eqnarray}
{h}(\bi{p})&=& 2t
\cos\gamma( p_x \pm p_y)\sigma_x+ 2t \sin\gamma ( p_x \mp p_y)\sigma_y, \label{lh}
\end{eqnarray}
where $\bi{p}=\bi{k}-X_i$ with $i=1,2$;  the two  signs indicate
the linearized Hamiltonian around the two distinct touched points  $X_1$ and $X_2$, respectively. The
Hamiltonian (\ref{lh}) has the form like   $h(\bi{p})=\sum_{ij}v_{ij}p_i\sigma_j$, based on which, though the $\gamma_5$ matrix is absent in $(2+1)$-dimensions,  we can define the chirality    for two-dimensional massless relativistic fermions as $
w=\textrm{sgn}[\det(v_{ij})]=\pm 1
$. We can regard the chiral relativistic fermions as two-dimensional Weyl fermions.
 From Eq.(\ref{lh}) and the definition of the chirality,   we have $w=-1, +1$ for Weyl nodes located at the degenerate points $X_1$ and $X_2$, respectively. It turns out that $X_1$ and $X_2$
have opposite chirality.

The Weyl nodes can also be interpreted as topological defects, i.e. vortices,  of the
planar vector field in momentum space like
$\bi{h}= (h_x,h_y)$
with $h_x=-2t\cos\gamma (\cos k_x+\cos k_y)$ and $h_y=-2t\sin\gamma(\cos k_x-\cos k_y)$,
which is defined by using the coefficients of  Pauli matrices in
Eq.(\ref{BH0}). The corresponding topological invariant is  the
winding number defined as  \cite{Sun}
\begin{eqnarray}
 w=\oint_{\cal C}\frac{d{\bi{k}}}{2\pi}
\cdot\left(\hat{h}_x\nabla\hat{h}_y
-\hat{h}_y\nabla\hat{h}_x\right),
\label{wn}
\end{eqnarray}
where $\hat{h}_i={h_i}/{|{\bi{h}}|}$.
For the definition Eq.(\ref{wn}), we calculate that the winding number of the  vortices at the degenerate points is $+ 1$ or $-1$, which is
consistent with the chirality defined above.

\em Symmetry protection.\em---The model in the first parameter range, i.e.,  $0<\gamma<\pi/2$, $t\neq 0$, $t_1=0$ and $v=0$, has a hidden symmetry, which is a discrete symmetry with an antiunitary compositor operator.
We will show that this hidden symmetry supports the existence of isolated band degeneracy.
 The corresponding composite symmetry operator can be written as,
\begin{eqnarray}
\Upsilon= \sigma_x KT_{\hat{x}}, \label{sym}
\end{eqnarray}
where $T_{\hat{x}}$ is a translation operator
which moves the lattice by $\hat{x}$  along the $x$ direction;
$K$ is the complex conjugation operator, which can also be regarded as a time-reversal operator for spinless particles; $\sigma_x$ is the Pauli
matrix representing the sublattice exchange.
 The model considered here is invariant under this composite transformation, i.e.,
$H=\Upsilon H\Upsilon^{-1}$, where
the inverse operator is $
\Upsilon^{-1}=\sigma_x KT_{\hat{x}}^{-1}$.
 The $\Upsilon$ operator has the character as
$\Upsilon^2=T_{2\hat{x}}$.

The Bloch function has the form $\Psi_{\bi{k}}(\bi{r})=[u^{(1)}(\bi{r}), u^{(2)}(\bi{r})]^T e^{i\bi{k}\cdot\bi{r}}$.
The symmetry operator $\Upsilon$ acts on the Bloch function
as follows
\begin{eqnarray}
\Upsilon\Psi_{\bi{k}}(\bi{r})&=&\left(\matrix{u^{(2)*}_{\bi{k}}(\bi{r}-\hat{x})e^{ik_x}\cr
u^{(1)*}_{\bi{k}}(\bi{r}-\hat{x})e^{ik_x}}\right)e^{-i\bi{k}\cdot\bi{r}}
=\Psi'_{\bi{k}'}(\bi{r}).\label{Bloch2}
\end{eqnarray}
 Because $\Upsilon$  is the symmetry operator of the system,
$\Psi'_{\bi{k}'}(\bi{r})$ must be a Bloch wave function of the
system. Thus, we obtain $\bi{k}'=-\bi{k}$,
$u^{(1)}_{\bi{k}'}(\bi{r})=u^{(2)*}_{\bi{k}}(\bi{r}-\hat{x})e^{ik_x}$
and
$u^{(2)}_{\bi{k}'}(\bi{r})=u^{(1)*}_{\bi{k}}(\bi{r}-\hat{x})e^{ik_x}$.
From Eq.(\ref{Bloch2}), it is easy to show that
the operator $\Upsilon$ has the effect when acting on  wave vectors
 as  $ \Upsilon: \bi{k}\rightarrow -\bi{k}=\bi{k}'$.
If $\bi{k}'=\bi{k}+\bi{K}_m$, where $\bi{K}_m$ is the reciprocal lattice vector, then we can say that $\bi{k}$ is a $\Upsilon$-invariant point in momentum space. In the Brillouin zone, the $\Upsilon$-invariant points  are the  points $\Gamma, M$ and $X_{1,2}$, which
are marked by green balls in Fig.\ref{fig1}(b). Suppose that $\bi{G}$ represents a $\Upsilon$-invariant point, then we have
$\Upsilon\Psi_{\bi{G}}(\bi{r})=\Psi'_{\bi{G}}(\bi{r})$.
Because of the equation $\Upsilon H\Upsilon^{-1}=H$, $\Psi_{\bi{G}}(\bi{r})$ and  $\Psi'_{\bi{G}}(\bi{r})$ are
both the eigenstates of Hamiltonian $H$ and  have the same
eigenenergy $E(\bi{G})$.  After the symmetry operator acts on the Bloch function two times, we have $
\Upsilon^2\Psi_{\bi{G}}(\bi{r})=T_{2\hat{x}}\Psi_{\bi{G}}(\bi{r})=e^{-2iG_{x}
}\Psi_{\bi{G}}(\bi{r})$,
where $G_{x}$ is the $x$ component of $\bi{G}$. From the above equation, we obtain $\Upsilon^2=e^{-2iG_{x}
}$, which is a function of the wave vector.  Substituting the wave vectors of the $\Upsilon$-invariant points $\Gamma, M$ and $X_{1,2}$, we obtain   $\Upsilon^2=1$ at $\Gamma, M$ and $\Upsilon^2=-1$ at $X_{1,2}$.  If a system is invariant under the action of an antiunitary operator and the  square of the operator is not equal to 1, there must be degeneracy protected by this antiunitary operator\cite{SM}.  Thus, the bands must be degenerate at the points $X_{1,2}$
 in the Brillouin zone, which is consistent with the Weyl nodes obtained from dispersion relation.

\em Quantum anomalous Hall effect.\em---
 For the second parameter range, i.e.,  $0<\gamma<\pi/2$, $t\neq 0$, and at least one of $t_1$ and $v$ being nonzero, there are nonvanishing  $H_1$ or $H_2$  in the total Hamiltonian. It is easy to show that $H_1$ and $H_2$ violate the hidden symmetry, i.e. $\Upsilon H_{1,2}\Upsilon^{-1}\neq H_{1,2}$. Thus, both of them can remove the degeneracy at Weyl nodes and  open a gap between the valence and conduction bands, turning the system into a insulator.

\begin{figure}[ht]
\includegraphics[width=0.47\columnwidth]{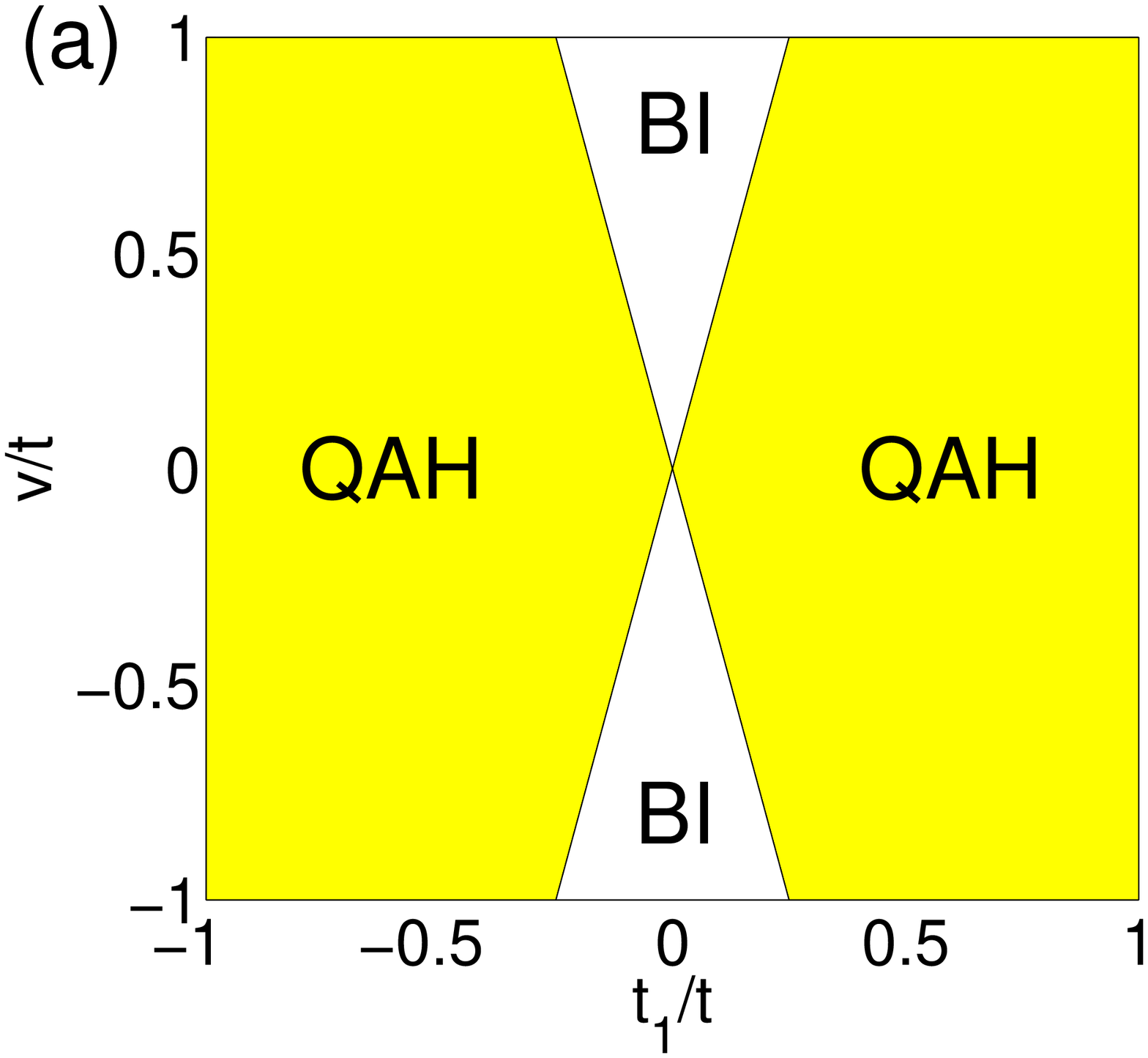}
\includegraphics[width=0.47\columnwidth]{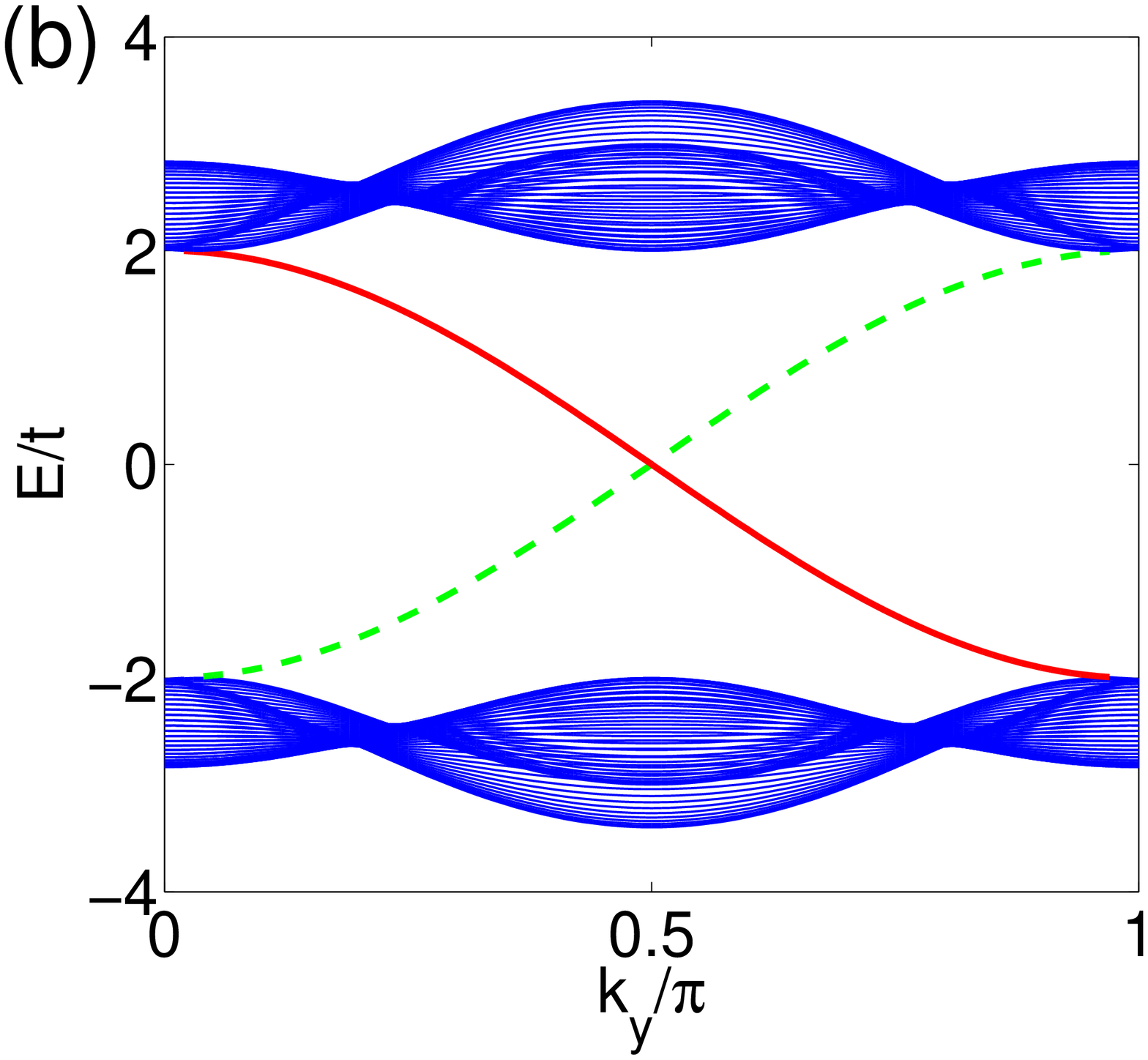}
\caption{(Color online). (a) Phase diagram for insulators. Here, QAH and BI represent quantum anomalous Hall insulator and trivial band insulator, respectively. (b) The edge states of the quantum anomalous Hall phase for $\gamma=\pi/4$, $t_1=0.8 t$ and $v=0.2t$. Red solid line and green dashed line represent edge states at two opposite edges, respectively.   }\label{fig3}
\end{figure}

The corresponding Bloch Hamiltonian can be written as
\begin{eqnarray}
{\cal H}(\bi{k})&=&-\cos\gamma \Omega_+\sigma_x
 - \sin\gamma\Omega_-\sigma_y
 +m\sigma_z,
 \label{BH2}
\end{eqnarray}
where $m=v-2t_1[\cos(k_x+k_y)-\cos(k_x-k_y)]$. Comparing with the first parameter range, there is an additive mass term.
Diagonalizing  the above equation, we obtain $E({\bi{k}}) =
  \pm  \sqrt{\cos^2\gamma\Omega_+^2+\sin^2\gamma\Omega_-^2+m^2}
$. Figure \ref{fig2}(b) shows that a gap between the conduction and valence bands is opened by the mass term.

 For an insulator, we can always characterize the occupied bands with the first Chern number calculated based on Berry phases in the first Brillouin zone.
The corresponding first Chern number can be defined as
$C_1=\frac{1}{4\pi}\int_{BZ} dk_x dk_y \hat{\bi{h}}\cdot\left( {\partial \hat{\bi{h}}}/{\partial k_x}\times  {\partial \hat{\bi{h}}}/{\partial k_y}\right)
$,
where $\hat{\bi{h}}=\bi{h}/|\bi{h}|$ with $\bi{h}$ being the coefficients of the Pauli matrices in Eq.(\ref{BH2}).
For quantum anomalous Hall states, the first Chern number must not be zero.
This is just the case when the mass term at the two distinct degenerate points has opposite sign, and the system has a nontrivial first Chern number $C_1=\pm 1$\cite{Haldane}.
The vanishing mass at one of the two distinct degenerate points $X_1$ and $X_2$, i.e., $v=\pm 4t_1$,  defines the boundary between quantum anomalous Hall insulator and
trivial band insulator in the phase diagram as shown in Fig.\ref{fig3}(a).
In the regime of the quantum anomalous Hall insulator, chiral edge states at each edge appear, which are shown  in Fig.\ref{fig3}(b).

At the boundary between the quantum anomalous Hall insulator and trivial band insulator, i.e., $v=\pm4t_1$, one of the two distinct Weyl nodes in the Brillouin zone reappears, as shown in Fig.\ref{fig2}(c). This can be explained by the recovery of the hidden symmetry $\Upsilon$ at some points in momentum space. For instance,
when $v=-4t_1$ is satisfied, the mass term in Eq.(\ref{BH2}) vanishes at  $X_1$, while it still exists at $X_2$. Thus, the symmetry $\Upsilon$ is recovered at $X_1$ point, i.e. $\Upsilon{\cal H}(X_1)\Upsilon^{-1}={\cal H}(X_1)$. Then, the Weyl node reappears at $X_1$ but does not at $X_2 $.  Similarly, for $v=4t_1$, the symmetry is recovered at $X_2$, but not at $X_1$, i.e.,  $\Upsilon{\cal H}(X_2)\Upsilon^{-1}={\cal H}(X_2)$ , then the Weyl node reoccurs at $X_2$ but not at $X_1$.

\em $2\pi$-flux topological semimetals.\em---
Now, we consider the third parameter range,  i.e., $\gamma=0$ and $v=0$ for the  total Hamiltonian $H$.
For this special case, the Bloch Hamiltonian can be written as,
\begin{eqnarray}
{\cal H}(\bi{k})&=&-\Omega_+\sigma_x -\Omega_z\sigma_z,\label{BH3}
\end{eqnarray}
where  $\Omega_z=2t_1[\cos(k_x+k_y)-\cos(k_x-k_y)]$.
Diagonalizing the Hamiltonian, we obtain the dispersion relation as
$E(\bi{k})=\pm \sqrt{\Omega_+^2+\Omega_z^2}$. The conduction and valence bands are touched at  point $M$ in the Brillouin zone as shown in Fig.\ref{fig2}(d). Around the degenerate point, the bands have a quadratic dispersion relation. For instance, retaining only the lowest terms around the degenerate point $M$, i.e., $(\pi,0)$, the Bloch Hamiltonian (\ref{BH3}) can be  written as,
\begin{eqnarray}
{\cal H}(\bi{k})&=&-t(p_x^2-p_y^2)\sigma_x-4t_1p_xp_y\sigma_z,
\label{BH30}
\end{eqnarray}
 where $\bi{p}=\bi{k}-M$. For each band crossing, from the formula (\ref{wn}), we can calculate a winding number with value of $+2$ or $-2$. These band crossings  correspond to vortices with
$2\pi$ or $-2\pi$ Berry flux and can also be regarded as a double-Weyl node consisting of two Weyl nodes with the same chirality, which is similar to the topological semimetal protected by $D_4$ point-group symmetry\cite{Sun}.

The $2\pi$-flux nodes in this parameter range can be interpreted by the protection of symmetry.
 In this parameter range, the model has a new composite symmetry such that $\Upsilon' H{\Upsilon'}^{-1}=H$ with
\begin{eqnarray}
\Upsilon'=\sigma_x KC_4T_{\hat{x}},
 \end{eqnarray}
 where $C_4$ is a rotation operation by $\pi/2$ around the normal vector of the $x-y$ plane. This composite antiunitary operator $\Upsilon'$ has  a more $C_4$ rotation  than the operator $\Upsilon$ and has the property that ${\Upsilon'}^{2}=C_2T_{\bi{a}_1}$ with $C_2=C_4^2$.
The action of symmetry operator $\Upsilon'$ on wave vectors has the effect that $\Upsilon':(k_x,k_y)\rightarrow(-k_y,k_x)$.
Thus, under this symmetry operation, the Bloch Hamiltonian is transformed as $\Upsilon'{\cal H}(k_x,k_y){\Upsilon'}^{-1}
={\cal H}(-k_y,k_x)$. It is very easy to show that the Bloch Hamiltonian is invariant at the point $M=(\pi,0)$
and   point $\Gamma=(0,0)$ in the   Brillouin zone.
  It is easy to show that $\Upsilon'^2=C_2T_{\bi{a}_1}=-1$ at   point $M$, while it is equal to  $1$ at  point $\Gamma$.
Therefore, the bands must be degenerate at  point $M$ in the
Brillouin zone, which is consistent with the dispersion relation calculated above.

The symmetry $\Upsilon'$ also guarantees that the dispersion relation is quadratic instead of linear near the degenerate point.
 The corresponding  Bloch Hamiltonian can be written in the form ${\cal H}(k_x,k_y)=f(k_x,k_y)\sigma_x+g(k_x,k_y)\sigma_z$, where $f(k_x, k_y)$ and $g(k_x,k_y)$ are functions of $k_x$ and $k_y$.   On  one hand, we have ${\cal H}(-k_y, k_x)=f(-k_y, k_x)\sigma_x+g(-k_y, k_x)\sigma_z$, on the other hand, $\Upsilon'{\cal H}(k_x, k_y){\Upsilon'}^{-1}=f(k_x,k_y)\sigma_x-g(k_x,k_y)\sigma_z$. Therefore, due to
 $\Upsilon'{\cal H}(k_x, k_y){\Upsilon'}^{-1}={\cal H}(-k_y, k_x)$, we obtain the equations $f(k_x,k_y)=f(-k_y, k_x)$ and $g(k_x,k_y)=-g(-k_y, k_x)$.
Since the degenerate point is located at $M$ in the Brillouin zone, the functions $f(k_x, k_y)$ and $g(k_x, k_y)$ can be expanded at point $M$ in   Taylor series as $f(k_x, k_y)=f( {M})+\sum_{i} c_i p_i+\frac{1}{2}\sum_{ij}c_{ij}p_ip_j+\cdots$ and $g(k_x, k_y)=g( {M})+\sum_{i} d_i p_i+\frac{1}{2}\sum_{ij}d_{ij}p_ip_j+\cdots$. Considering the above equations satisfied by $f(k_x, k_y)$ and $g(k_x, k_y)$, it is easy to show that all the first order terms vanish for $f(k_x, k_y)$ and $g(k_x, k_y)$, and the lowest order nonvanishing terms are the second order terms, which is consistent with Eq.(\ref{BH30}). Therefore, the fact that the dispersion relation is quadratic  near the degenerate point can be interpreted from the protection of symmetry.
More general and detailed proof is presented in Supplemental Material\cite{SM}.

\em{Experimental techniques for physical realization}\em--The high controllability and large number of detection techniques  of cold atoms in optical lattices make them a  platform we can use to realize many models in condensed matter physics.
The model considered by us can be realized by applying $^{40}{\rm K}$ cold atoms trapped in spin-dependent optical lattice\cite{Bloch}. The accompanying phase of hopping can be realized by laser-induced gauge potentials\cite{Hou,Lin,Aidelsburger}.  The interferometric approach proposed by Abanian et al.\cite{Abanin} can be used to detect the winding number at degenerate points of topological semimetals and the Chern number of the quantum anomalous Hall insulator. The Chern number of insulators can also be measured with the time-of-flight method proposed by Alba et al.\cite{Alba} and the method of measuring Bloch eigenstates at   symmetric points of the Brillouin zone proposed by Liu et al.\cite{Liuxj}.

   \em Conclusion.\em---In summary,  we have shown that, besides  time-reversal symmetry and particle-hole symmetry, there exists a class of discrete symmetries with antiunitary composite operators contributing  to  the protection of degeneracies in two-dimensional systems.
   In order to clearly manifest the close relation between this kind  of hidden symmetries and degeneracies in two-dimensional systems, we  have studied a  fermionic square lattice for example. This model supports the existence of a two-dimensional Weyl semimetal, quantum anomalous Hall effect, and $2\pi$-flux topological semimetal in different parameter ranges.  We have   shown that the two-dimensional Weyl semimetal and $2\pi$-flux topological semimetal are protected by two distinct hidden symmetries, respectively. When these hidden symmetries are broken, a gap opens between the conduction and valence bands and quantum anomalous Hall effect appears in the  appropriate parameters. We also found that the part of Weyl nodes reoccur when the parameters approach to the boundary between quantum anomalous Hall insulator and trivial band  insulator in the phase diagram, and they are also protected by the hidden symmetry.

We thank W. Chen, X. Wan, and X. J. Liu for helpful discussions.
 This work was  supported by the National Natural Science Foundation
of China under Grants No. 11004028 and No. 11274061.

\newpage
\setcounter{figure}{0}
\renewcommand{\thefigure}{S\arabic{figure}}
\setcounter{equation}{0}
\renewcommand{\theequation}{S\arabic{equation}}
\setcounter{table}{0}
\renewcommand{\thetable}{S\arabic{table}}

\begin{widetext}
\begin{center}
 {\textbf{ Supplementary Material }} 
 \end{center}

\subsection{A. Proof of the protection of degeneracy by an anti-unitary operator}

We assume that $\Upsilon$ is an anti-unitary operator. $H$ is a Hamiltonian and satisfies $[H, \Upsilon]=0$, i.e., $H$ is $\Upsilon$-invariant.
$\Upsilon^2$ must be a unitary operator and satisfies the equation $[H, \Upsilon^2]=0$. Therefore, the Hamiltonian $H$ and $\Upsilon^2$ have common eigenstates.    Suppose that $|\Psi\rangle$ is a common eigenstate of the Hamiltonian $H$ and $\Upsilon^2$ and obey the equations $\Upsilon^2|\Psi\rangle=\alpha|\Psi\rangle$ and $H|\Psi\rangle=E_0|\Psi\rangle$. Then,
due to $[H, \Upsilon]=0$,  $|\Psi'\rangle\equiv \Upsilon|\Psi\rangle$ is also a eigenstate of Hamiltonian $H$ with the eigenenergy $E_0$. We take the inner product of the states $|\Psi'\rangle$ and $|\Psi\rangle$ as follows,
\begin{eqnarray}
\langle \Psi'|\Psi\rangle=\langle \Upsilon\Psi|\Upsilon \Psi'\rangle=\langle \Psi'|\Upsilon^2\Psi\rangle =\alpha\langle \Psi'|\Psi\rangle,
\end{eqnarray}
where we have used the property that $\langle \Upsilon \varphi|\Upsilon \xi\rangle=\langle\xi|\varphi\rangle$ for the anti-unitary operator $\Upsilon$.  Thus, we have
\begin{eqnarray}
(1-\alpha)\langle \Psi'|\Psi\rangle=0
\end{eqnarray}
If $\alpha\neq 1$, $\langle \Psi'|\Psi\rangle=0$ must be satisfied.  Namely, $|\Psi'\rangle$ and $\Psi\rangle$ are  orthogonal to each other. On the other hand, $|\Psi'\rangle$ and $\Psi\rangle$ have the same eigenenergy $E_0$. Thus,  the system is degenerate. In summary, we conclude that \em if a system is invariant under the action of an anti-unitary operator and the  square of the operator is not equal to 1, there must be degeneracy protected by this anti-unitary operator.\em     This proof is similar to the proof of Kramers degeneracy protected by time-reversal symmetry in many textbooks.

\subsection{B. Proof of the quadratic dispersion for the third parameter range}

The
corresponding Bloch functions for the $A$ and $B$ sublattices  can
be written, respectively, as\cite{Bena}
\begin{eqnarray}
&&\psi^A_{\bi{k}}(\bi{r})=\frac{1}{\sqrt{N}}\sum_{i
 }e^{i\bi{k}\cdot\bi{R}^A_i}w^A_i(\bi{r}-\bi{R}^A_i),\label{basis1}\\
&&\psi^B_{\bi{k}}(\bi{r} )=\frac{1}{\sqrt{N}}\sum_{i
}e^{i\bi{k}\cdot\bi{R}^B_i}w^B_i(\bi{r}-\bi{R}^B_i)\label{basis2}
\end{eqnarray}
 where $N$ is  the number of lattice sites in each sublattice, and $\bi{R}^{B}_i=\bi{R}^A_i+\hat{x}$ with $\hat{x}=(1,0)$.
In the main text, the Bloch Hamiltonian has the form as follows,
\begin{eqnarray}
{\cal H}(\bi{k})=f(\bi{k})\sigma_x +g(\bi{k})\sigma_z
\label{Bloch}
\end{eqnarray}
$\Upsilon'=\sigma_x KC_4T_{(\bi{a}_1+\bi{a}_2)/2}$ is the symmetry operator, which acts on the Bloch Hamiltonian and satisfies the following equation\cite{Fang},
\begin{eqnarray}
\Upsilon'{\cal H}(\bi{k}){\Upsilon'}^{-1}={\cal H}(\Upsilon' \bi{k})
\label{trans}
\end{eqnarray}
Substituting the symmetry operator $\Upsilon'$ and the Bloch Hamiltonian  (\ref{Bloch}) into  the left-hand side of Eq.(\ref{trans}), we obtain
\begin{eqnarray}
\Upsilon'{\cal H}(k_x, k_y){\Upsilon'}^{-1}=f(k_x, k_y)\sigma_x-g(k_x, k_y)\sigma_z
\label{left}
\end{eqnarray}
From the main text, we know that $\Upsilon': (k_x, k_y)\rightarrow (-k_y, k_x)$. Thus the right-hand side of Eq.(\ref{trans}) can be written as
\begin{eqnarray}
{\cal H}(-k_y,k_x)=f(-k_y,k_x)\sigma_x+g(-k_y, k_x)\sigma_z
\label{right}
\end{eqnarray}
Combining Eqs. (\ref{trans}), (\ref{left}) and (\ref{right}), we arrive at the following constrained conditions,
\begin{eqnarray}
&&f'(k_x,k_y)\equiv f(-k_y, k_x)=f(k_x, k_y),\\
&&g'(k_x,k_y)\equiv  g(-k_y, k_x)=-g(k_x, k_y)
\end{eqnarray}
Since $M$ is the degenerate point in Brillouin zone,  the functions $f(k_x, k_y)$ and $f'(k_x, k_y)$ can be expanded in Taylor series as follow
\begin{eqnarray}
f(k_x,k_y)=f(M_x, M_y)+\left.\frac{\partial f}{\partial k_x}\right|_\bi{M}p_x+\left.\frac{\partial f}{\partial k_y}\right|_\bi{M}p_y
+\frac{1}{2}\left.\frac{\partial^2 f}{\partial k_x^2}\right|_\bi{M}p_x^2+\frac{1}{2}\left.\frac{\partial^2 f}{\partial k_y^2}\right|_\bi{M}p_y^2
+ \left.\frac{\partial^2 f}{\partial k_x\partial k_y}\right|_\bi{M}p_xp_y+\cdots
\label{Taylor1}
\end{eqnarray}
and
\begin{eqnarray}
f'(k_x, k_y)=f'(M_x, M_y)+\left.\frac{\partial f'}{\partial k_x}\right|_\bi{M}p_x+\left.\frac{\partial f'}{\partial k_y}\right|_\bi{M}p_y
+\frac{1}{2}\left.\frac{\partial^2 f'}{\partial k_x^2}\right|_\bi{M}p_x^2+\frac{1}{2}\left.\frac{\partial^2 f'}{\partial k_y^2}\right|_\bi{M}p_y^2
+ \left.\frac{\partial^2 f'}{\partial k_x\partial k_y}\right|_\bi{M}p_xp_y+\cdots
\label{Taylor2}
\end{eqnarray}
where $p_x=k_x-M_x, p_y=k_y-M_y$. Comparing the first order terms of the two Taylor series (\ref{Taylor1}) and (\ref{Taylor2}), we obtain
\begin{eqnarray}
&&\left.\frac{\partial f'}{\partial k_x}\right|_\bi{M}=\left.\frac{\partial f}{\partial k_x}\right|_\bi{M}, \ \ \ \ \left.\frac{\partial f'}{\partial k_y}\right|_\bi{M}=\left.\frac{\partial f}{\partial k_y}\right|_\bi{M} \label{EG1}
\end{eqnarray}
Substituting the definition $f'(k_x,k_y)\equiv f(-k_y, k_x)$ into Taylor series (\ref{Taylor2}), we have,
\begin{eqnarray}
f'(k_x, k_y)&=&f(-k_y,k_x)\nonumber\\
&=&f(M'_x, M'_y)+\left.\frac{\partial f}{\partial k_y}\right|_{\bi{M}'}p_x-\left.\frac{\partial f}{\partial k_x}\right|_{\bi{M}'}p_y
+\frac{1}{2}\left.\frac{\partial^2 f}{\partial k_y^2}\right|_{\bi{M}'}p_x^2+\frac{1}{2}\left.\frac{\partial^2 f}{\partial k_y^2}\right|_{\bi{M}'}p_y^2
- \left.\frac{\partial^2 f}{\partial k_x\partial k_y}\right|_{\bi{M}'}p_xp_y+\cdots
\end{eqnarray}
where $M'=\Upsilon M$, i.e., $(M'_x, M'_y)=(-M_y, M_x)$. Then, we arrive at
\begin{eqnarray}
&&\left.\frac{\partial f'}{\partial k_x}\right|_\bi{M}=\left.\frac{\partial f}{\partial k_y}\right|_{\bi{M}'}, \ \ \ \ \left.\frac{\partial f'}{\partial k_y}\right|_\bi{M}=-\left.\frac{\partial f}{\partial k_x}\right|_{\bi{M}'}\label{aa}
\end{eqnarray}
We choose the wave function (\ref{basis1}) and (\ref{basis2}) as the basis and there is a phase difference between the wave function due to $\bi{R}^B_i=\bi{R}^A_i+\hat{x}$, so the off-diagonal elements of the Bloch Hamiltonian includes $e^{ik_x}$,  $e^{-ik_x}$  factors or mixture of them. From the main text, we know that
$M'_x-M_x=\pm \pi$, so we can arrive at $f(\bi{k}-\bi{M}'+\bi{M})=-f(\bi{k})$. Substituting the relation into  Eq.(\ref{aa}), we obtain
\begin{eqnarray}
&&\left.\frac{\partial f'}{\partial k_x}\right|_\bi{M}=-\left.\frac{\partial f}{\partial k_y}\right|_{\bi{M}}, \ \ \ \ \left.\frac{\partial f'}{\partial k_y}\right|_\bi{M}=\left.\frac{\partial f}{\partial k_x}\right|_{\bi{M}}\label{EG2}
\end{eqnarray}
Solving the equation group (\ref{EG1}) and (\ref{EG2}), we obtain the solution
\begin{eqnarray}
\left.\frac{\partial f}{\partial k_x}\right|_{\bi{M}}=\left.\frac{\partial f}{\partial k_y}\right|_\bi{M}=0\label{first1}
\end{eqnarray}

Similarly, $g(k_x, k_y)$ can also be expanded in Taylor series around the degenerate points $\bi{M}$ as
\begin{eqnarray}
g(k_x,k_y)=g(M_x, M_y)+\left.\frac{\partial g}{\partial k_x}\right|_\bi{M}p_x+\left.\frac{\partial g}{\partial k_y}\right|_\bi{M}p_y
+\frac{1}{2}\left.\frac{\partial^2 g}{\partial k_x^2}\right|_\bi{M}p_x^2+\frac{1}{2}\left.\frac{\partial^2 g}{\partial k_y^2}\right|_\bi{M}p_y^2
+ \left.\frac{\partial^2 g}{\partial k_x\partial k_y}\right|_\bi{M}p_xp_y+\cdots
\end{eqnarray}
Following the similar process as the function$f(k_x, k_y)$, we can also obtain the same results about $g(k_x, k_y)$ as
\begin{eqnarray}
\left.\frac{\partial g}{\partial k_x}\right|_{\bi{M}}=\left.\frac{\partial g}{\partial k_y}\right|_\bi{M}=0\label{first2}
\end{eqnarray}
From Eqs.(\ref{first1}) and (\ref{first2}), we know that the linear terms in the Bloch Hamiltonian vanish. It is easy to show that the lowest non-zero terms in the Taylor series of the Bloch Hamiltonian are the second order terms. Therefore, we conclude that around the degenerate point, the Bloch Hamiltonian and the dispersion relation is quadratic.

\subsection{C. Other two-dimensional lattices with hidden symmetry}

\begin{figure}[ht]
\includegraphics[width=0.35\columnwidth]{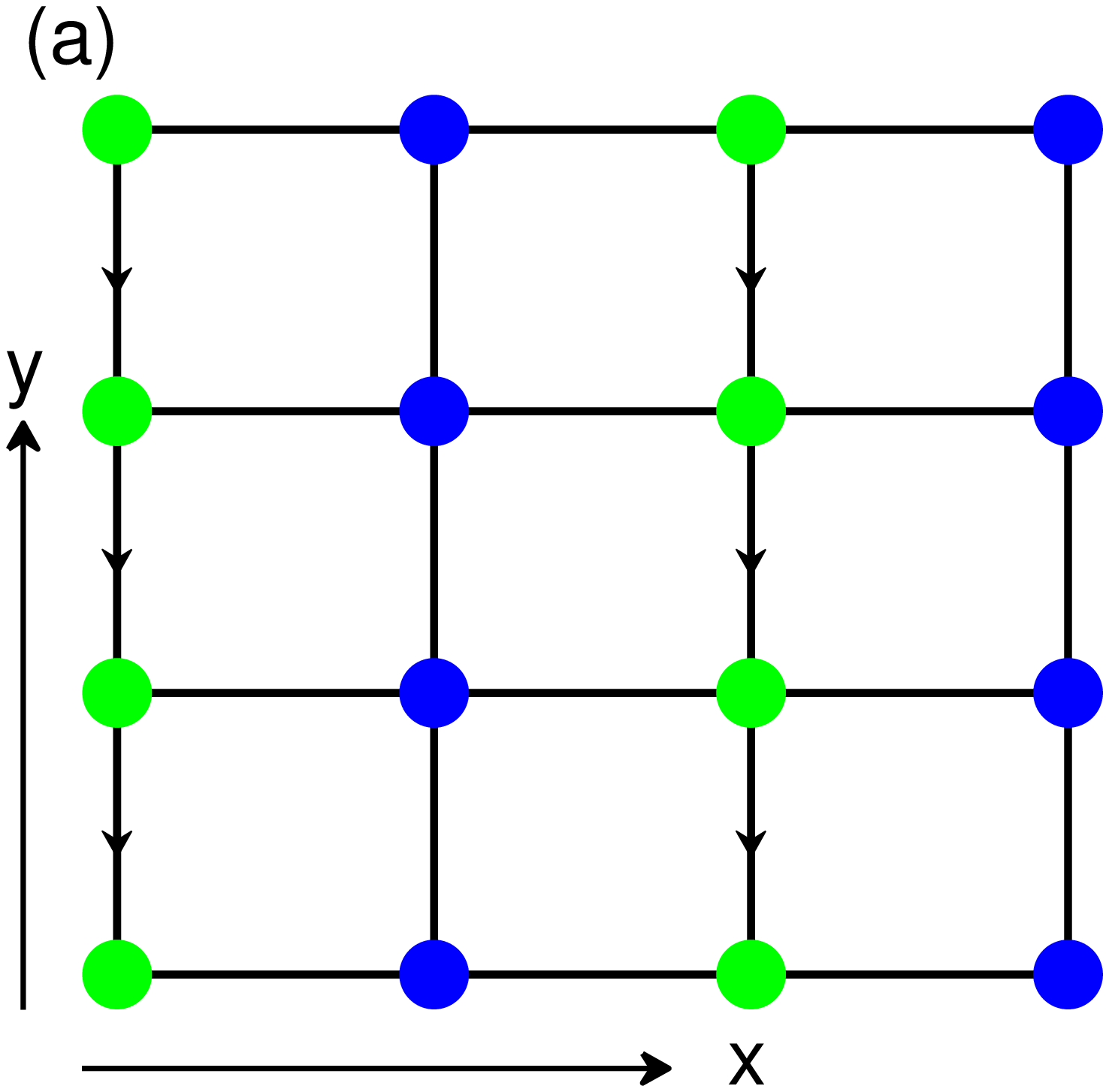}
\includegraphics[width=0.35\columnwidth]{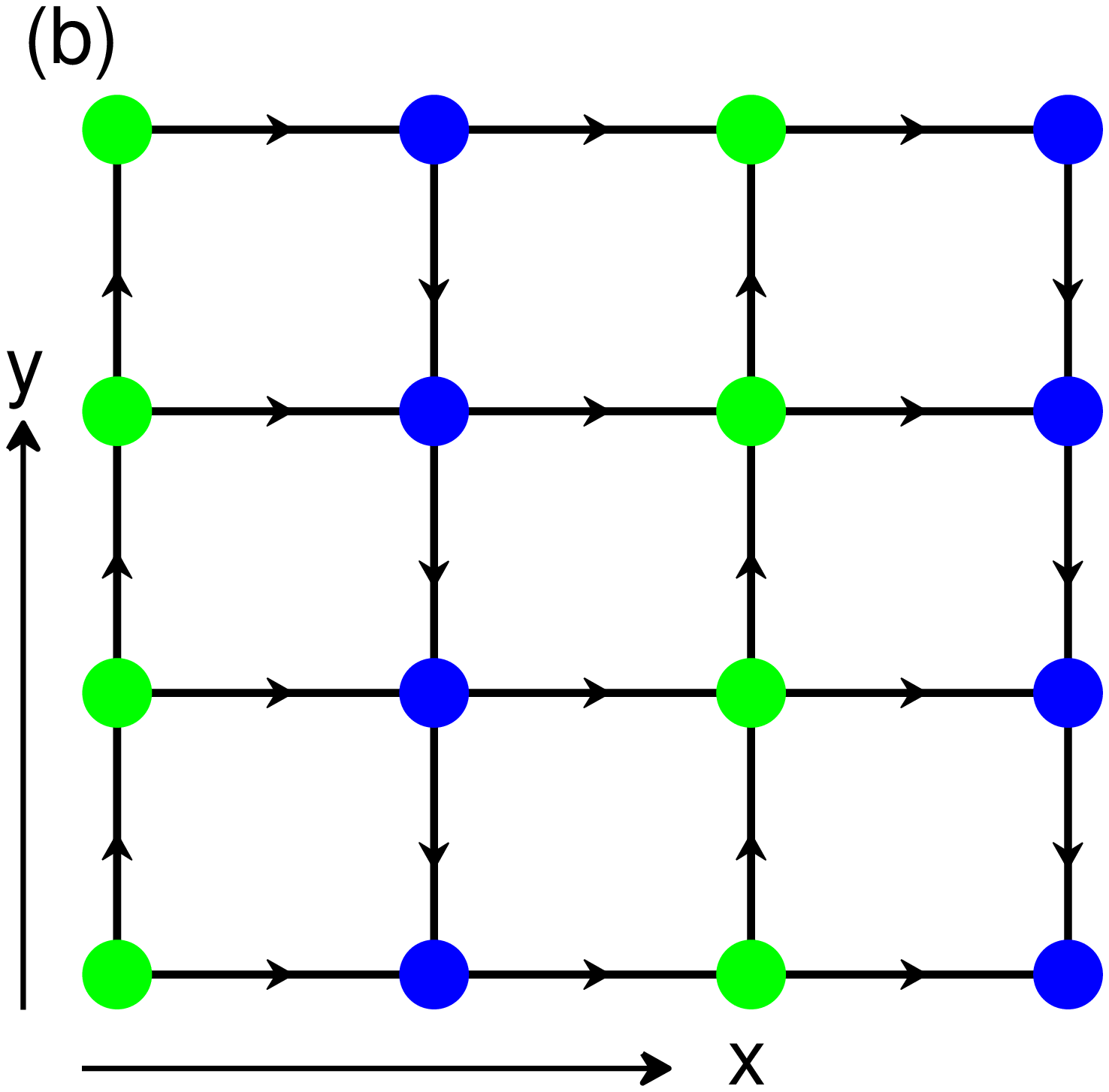}
\includegraphics[width=0.35\columnwidth]{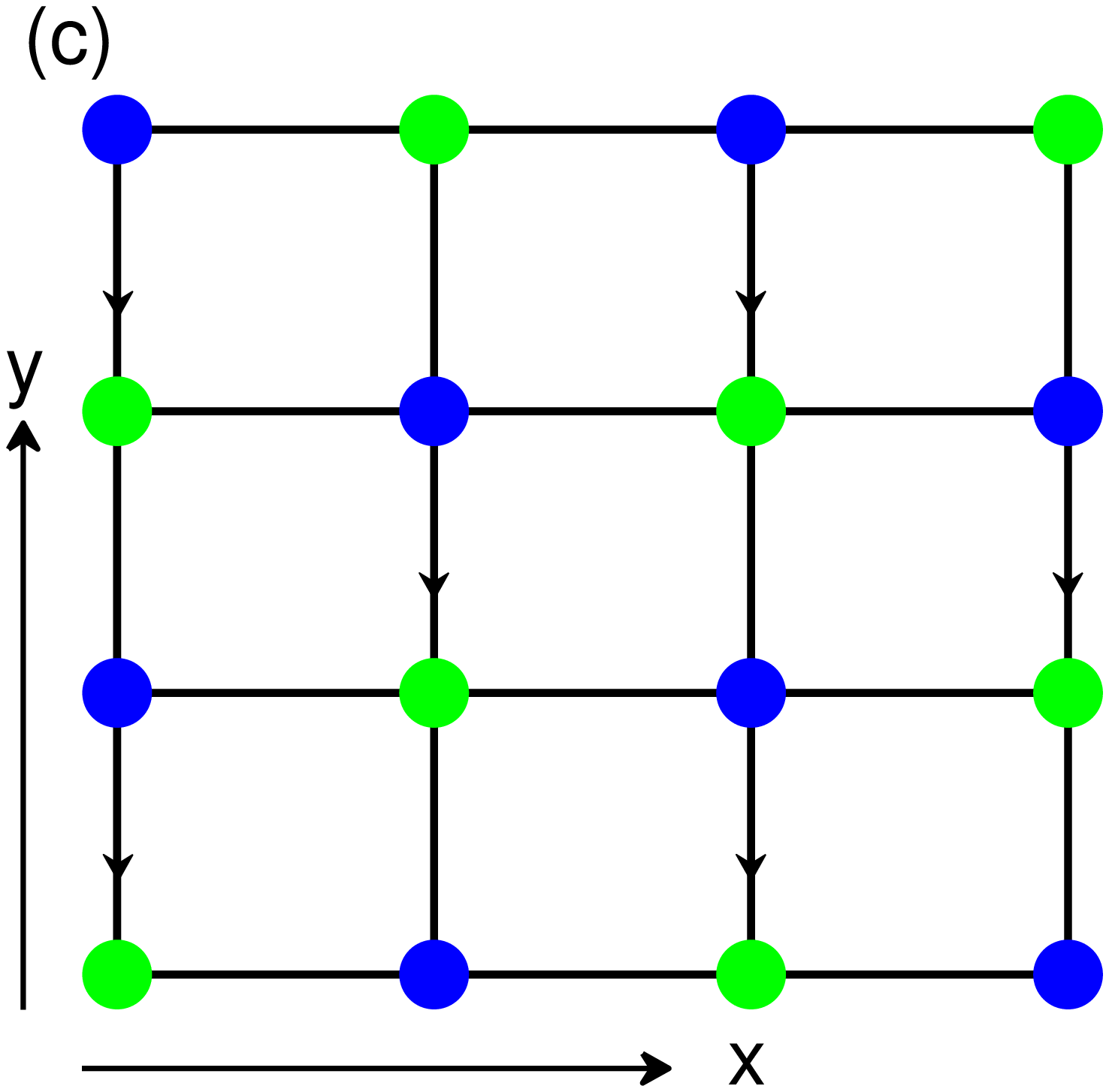}
\caption{  Schematic of the lattices. Here, the green  and blue
balls represent the lattice sites of sublattices $A$ and $B$,
respectively. The arrows represent the hopping-accompanying phase. (a) Model 1: the hopping-accompanying phase is $0<\gamma<2\pi$:   (b) Model 2: the hopping-accompanying phases are $0<\beta<\pi$ in the $x$ direction and $0<\gamma<\pi$ in the $y$ direction; and (c) Model 3: the hopping-accompanying phase is $0<\gamma<2\pi$. }\label{S_fig1}
\end{figure}

\begin{table}[ht]
\begin{center}
\caption{Summary of three models with hidden symmetry-protected degeneracy}\label{tab1}
\begin{tabular}{|c|p{2cm}|p{2cm}|p{3.9cm}|p{2.2cm}|p{2.6cm}|p{2.6cm}|}\hline
Model&Primitive lattice vectors&Primitive
reciprocal  lattice vectors&Symmetry operator&The effect of symmetry acting on wave vectors&Symmetry-invariant points&
Symmetry-protected degenerate points \\ \hline
1&$\bi{a}_1=(2,0)$, $\bi{a}_2=(0,1)$ &$\bi{b}_1=(\pi, 0)$,  $\bi{b}_2=(0, 2\pi)$ &$\Upsilon= (e^{i\gamma})^{i_y}\sigma_x K T_{\bi{a}_1/2}$, \ \ \ \ \ \ \ \ \ \ \ \ \ \ \
 $\Upsilon^2=T_{\bi{a}_1} $&
$\Upsilon: (k_x, k_y)\rightarrow(-k_x, \gamma-k_y)$&$(\pi/2, \gamma/2)$, $(\pi/2, \gamma/2-\pi)$,\ \ \ \ \ \ \ \ \ \ \ \ \ \ \ \ \ \ \ \ \ \ \ \ \ \ \ \ \ \ \ \ \ \ \ \ \ \ \ \ \ \ \ \ \ \ \ \ \ \
$(0, \gamma/2)$, $(0, \gamma/2-\pi)$
&$(\pi/2, \gamma/2)$, $(\pi/2, \gamma/2-\pi)$\\ \hline
2&$\bi{a}_1=(2,0)$, $\bi{a}_2=(0,1)$  &$\bi{b}_1=(\pi, 0)$,  $\bi{b}_2=(0, 2\pi)$ &$\Upsilon=(e^{2i\beta})^{i_x}\sigma_x K T_{\bi{a}_1/2}$,
 $\Upsilon^2=e^{2i\beta}T_{\bi{a}_1} $&$\Upsilon: (k_x, k_y)\rightarrow (2\beta-k_x, -k_y)$
&$(\beta-\pi/2, 0)$,  $(\beta-\pi/2, \pi)$, $(\beta, 0)$, $(\beta, \pi)$ &$(\beta-\pi/2, 0)$,\ \ \ \ \ \ \ \ \ \ \ \ \ \ \  $(\beta-\pi/2, \pi)$ \\ \hline
3&$\bi{a}_1=(1,1)$, $\bi{a}_2=(1,-1)$ &$\bi{b}_1=(\pi, \pi)$, $\bi{b}_2=(\pi, -\pi)$ &$\Upsilon= (e^{i\gamma})^{i_y}\sigma_x K T_{(\bi{a}_1+\bi{a}_2)/2}$,
 $\Upsilon^2=T_{\bi{a}_1+\bi{a}_2} $ & $\Upsilon: (k_x, k_y)\rightarrow(-k_x, \gamma-k_y)$ &$(-\pi/2, \gamma/2-\pi/2)$, $(\pi/2, \gamma/2-\pi/2)$, $(0, \gamma/2)$, $(0,\gamma/2-\pi)$ &$(-\pi/2, \gamma/2-\pi/2)$, $(\pi/2, \gamma/2-\pi/2)$ \\ \hline
\end{tabular}
\end{center}
 Note: Here, $i_x$ and $i_y$ are the $x$ and $y$  components of the coordinate $i$ of lattice sites, respectively. \ \ \ \ \ \ \ \ \ \ \ \ \ \ \ \ \ \ \ \ \ \ \ \ \
\end{table}

In order to    show the ubiquity of
 symmetry-protected isolated point degeneracies, we show three other models as shown in Fig.\ref{S_fig1} in this Supplementary Material.
  Here, for brevity,  we assume  the distance between the two neighbor lattice sites $d=1$. All the lattices considered here consist of two sublattices marked by blue and green balls in Fig.\ref{S_fig1}. In all the models, hopping between neighbor sites is allowed and the hopping amplitudes are the same. For some hoppings between neighbor sites, there exist hopping-accompanying phases, which are represented by arrows in Fig.\ref{S_fig1}. The magnitudes of hopping-accompanying phase represented by arrows are (i) $0<\gamma<2\pi$ in the $y$ direction for Model 1, (ii) $0<\beta<\pi$ in the $x$ direction, $0<\gamma<\pi$ in the $y$ direction for Model 2,
  (iii) $0<\gamma<2\pi$ in the $y$ direction for Model 3, respectively.
      The corresponding primitive lattice vectors, primitive reciprocal lattice vectors, symmetry operator, the transformation of wave vector under symmetry operation, symmetry-invariant points in the Brillouin zone, and symmetry-protected degenerate points are summarized in Table \ref{tab1}.

\end{widetext}


\begin{thebibliography}{99}
\bibitem{Hasan} M.Z. Hasan and C.L. Kane, Rev. Mod. Phys.
\textbf{82}, 3045 (2010). 

\bibitem{Qi} X.L. Qi and S.C. Zhang, Rev. Mod. Phys. \textbf{83},
1057 (2011). 

\bibitem{Schnyder}A.P. Schnyder, S. Ryu, A. Furusaki, and A.W.W
Ludwig, Phys. Rev. B \textbf{78}, 195125 (2008); \em ibid\em. AIP
Conf. Proc. \textbf{1134}, 10 (2009); A. Kitaev, AIP Conf. Proc.
\textbf{1134}, 22 (2009).


\bibitem{Thouless}D.J. Thouless, M. Kohmoto, M.P. Nightingale,
 M. den Nijs,  Phys. Rev. Lett.   \textbf{49},  405 (1982). 

\bibitem{Haldane} F.D.M. Haldane, Phys. Rev. Lett. \textbf{61}, 2015 (1988).

\bibitem{LiuQA} X.J. Liu, X. Liu, C. Wu, and J. Sinova, Phys. Rev. A \textbf{81}, 033622 (2010). 

\bibitem{Kane}  C.L. Kane,     E.J. Mele, Phys. Rev. Lett.
\textbf{95},  226801 (2005); \em ibid\em. \textbf{95},     146802 (2005). 

\bibitem{Read} N. Read and D. Green, Phys. Rev. B \textbf{61}, 10267
(2000).



\bibitem{Qi2} X.L. Qi, T.L. Hughes, S. Raghu, and S.C. Zhang, Phys.
Rev. Lett. \textbf{102}, 187001 (2009). 

\bibitem{Wan} X. Wan, A.M. Turner, A. Vishwanath, and S.Y. Savrasov,
Phys. Rev. B \textbf{83}, 205101 (2011). 

\bibitem{Xu} G. Xu, H. Weng, Z. Wang, X. Dai, and Z. Fang, Phys.
Rev. Lett. \textbf{107}, 186806 (2011). 

\bibitem{Burkov1} A.A. Burkov, M.D. Hook, and L. Balents, Phys. Rev.
B \textbf{84}, 235126 (2011). 

\bibitem{Burkov2} A.A. Burkov and L. Balents, Phys. Rev. Lett.
\textbf{107}, 127205 (2011). 

\bibitem{Fang} C. Fang, M.J. Gilbert, X. Dai, and B.A. Bernevig,
Phys. Rev. Lett. \textbf{108}, 266802 (2012). 

\bibitem{Zyuzin} A.A. Zyuzin, S. Wu, and A.A. Burkov, Phys. Rev.
\textbf{85}, 165110 (2012). 

\bibitem{Sun} K. Sun, W.V. Liu, A. Hemmerich and S. Das Sarma, Nat.
Phys. \textbf{8}, 67 (2012). 

\bibitem{Jiang} J.H. Jiang, Phys. Rev. A \textbf{85}, 033640 (2012).

\bibitem{Hosur} P. Hosur, S.A. Parameswaran, and A. Vishwanath,
Phys. Rev. Lett. \textbf{108}, 046602 (2012). 

\bibitem{Delplace} P. Delplace, J. Li, and D Carpentier, Europhys.
Lett. \textbf{97}, 67004 (2012). 



\bibitem{Herring} C. Herring, Phys. Rev. \textbf{52}, 365 (1937). 

\bibitem{Balents} L. Balents, Phys. \textbf{4}, 36 (2011). 

\bibitem{Lim} L.K. Lim, C. M. Smith, and A. Hemmerich, Phys. Rev. Lett. \textbf{100}, 130402 (2008).


\bibitem{Hou} J.M. Hou, W.X. Yang and X.J. Liu, Phys. Rev. A
\textbf{79}, 043621 (2009). 



\bibitem{Goldman}N. Goldman, A. Kubasiak, A. Bermudez, P. Gaspard, M. Lewenstein, and M. A. Martin-Delgado, Phys. Rev. Lett. \textbf{103}, 035301 (2009). 
\bibitem{Bercioux}D. Bercioux, D. F. Urban, H. Grabert, and W. H\"ausler, Phys. Rev. A \textbf{80}, 063603 (2009). 

\bibitem{Goldman2}N. Goldman, E. Anisimovas, F. Gerbier, P. \"Ohberg, I. B. Spielman, G.
Juzeli\={u}nas, New J. Phys. \textbf{15}, 013025 (2013).

\bibitem{SM} See Supplemental Material for details.

\bibitem{Bena} C. Bena and G. Montambaux, New J. Phys., \textbf{11}, 095003 (2009). 



\bibitem{Bloch} I. Bloch, J. Dalibard, and W. Zwerger, Rev. Mod. Phys. \textbf{80}, 885 (2008).

\bibitem{Lin} Y.J. Lin, R.L. Compton, K. Jim\'enez-Garc\'ia, J.V. Porto, and I.B. Spielman, Nature, \textbf{462}, 628 (2009).

\bibitem{Aidelsburger}M. Aidelsburger, M. Atala, S. Nascimb\`ene, S. Trotzky, Y.A. Chen, and I. Bloch, Phys. Rev. Lett. \textbf{107}, 255301 (2011).

\bibitem{Abanin} D.A. Abanin, T. Kitagawa, I. Bloch, and E. Demler,
Phys. Rev. Lett. \textbf{110}, 165304 (2013). 

\bibitem{Alba} E. Alba, X. Fernandez-Gonzalvo, J. Mur-Petit, J.K. Pachos, and J.J. Garcia-Ripoll,
Phys. Rev. Lett. \textbf{107}, 235301 (2011).

\bibitem{Liuxj} X.J. Liu,  K.T. Law,  T.K. Ng, and P.A. Lee, arXiv:1306.5223. 





\end{thebibliography}

\begin{thebibliography}{99}
\bibitem{Bena} C. Bena and G. Montambaux, New J. Phys. \textbf{11}, 095003 (2009).
\bibitem{Fang} C. Fang, M.J. Gilbert, X. Dai, and B.A. Bernevig,
Phys. Rev. Lett. \textbf{108}, 266802 (2012).
\end{thebibliography}
\end{document}